# Films of Bacteria at Interfaces (FBI): Remodeling of Fluid Interfaces by *Pseudomonas aeruginosa*.

Tagbo H. R. Niepa[1,#], Liana Vaccari[1,#], Robert L. Leheny[2], Mark Goulian[3], Daeyeon Lee[*,1] and Kathleen J. Stebe[*,1]

[1]Department of Chemical and Biomolecular Engineering, University of Pennsylvania, Philadelphia, PA 19104, USA

[2]Department of Physics and Astronomy, Johns Hopkins University, Baltimore, MD 21218, USA

[3]Department of Biology, University of Pennsylvania, Philadelphia, PA 19104, USA,

#Equal contribution authors

*Corresponding author:

Email: daeyeon@seas.upenn.edu, kstebe@seas.upenn.edu.

**Abstract:**

Bacteria at fluid interfaces endure physical and chemical stresses unique to these highly asymmetric environments. The responses of *Pseudomonas aeruginosa* PAO1 and PA14 to a hexadecane-water interface are compared. PAO1 cells form elastic films of bacteria, excreted polysaccharides and proteins, whereas PA14 cells move actively without forming an elastic film. Studies of PAO1 mutants show that, unlike solid-supported biofilms, elastic interfacial film formation occurs in the absence of flagella, pili, or certain polysaccharides. Highly induced genes identified in transcriptional profiling include those for putative enzymes and a carbohydrate metabolism enzyme, *alkB2*; this latter gene is not upregulated in PA14 cells. Notably, PAO1 mutants lacking the *alkB2* gene fail to form an elastic layer. Rather, they form an active film like that formed by PA14. These findings demonstrate that genetic expression is altered by interfacial confinement, and suggest that the ability to metabolize alkanes may play a role in elastic film formation at oil-water interfaces.

**INTRODUCTION**

Bacteria trapped at interfaces can form single and poly-species microbial communities. Examples include slimes on viscid mucus[1] in the lungs of cystic fibrosis patients[2], and pellicles that appear at interfaces of contaminated beer[3]. The formation of films of bacteria at interfaces impacts ecology, human health, and industries including pharmaceuticals, food production and oil recovery. The development and mechanics of interface-associated structures differ from those of classic surface-attached biofilms[4,5]. For example, fluid interfaces can trap microorganisms, precluding the reversible attachment that typically initiates solid surface associated biofilms, and physical forces on interface-trapped cells can trigger their aggregation and alter their subsequent development. While the mechanics of films of bacteria at interfaces have received some attention, little is known about their biological implications.

A microbe attached to a fluid interface with interfacial tension $\gamma$ eliminates a patch of interface of area $\Delta A$ and lowers the free energy by the amount $\gamma \Delta A$. This trapping energy must be overcome by work to detach the microbe from the interface. A typical bacterial cell body, with length and width between 1 – 10 µm, will experience trapping energies large enough to make bacteria attachment essentially irreversible. Furthermore, interfacial tension exerts a force along the contact line where the interface intersects the bacterium. Thus, bacteria trapped at interfaces are under physical tension. Furthermore, bacteria can change the shape of the interface, and could interact and assemble by capillary interactions to minimize the interfacial area[6]. In addition, fluid interfaces are chemically anisotropic; for example, at an alkane-water interface, the chemistry transitions from a polar, aqueous environment to a non-polar alkane environment over distances of a few molecular length scales. These and other physical and chemical factors unique to interfaces pose challenges to microbial survival. Some microbes respond to these physicochemical challenges by restructuring the interface to form viscoelastic interfacial films.[7] However, the generality of this response and its relationship to cell physiology are unknown, in particular at oil-water interfaces. A deeper understanding could advance the field of biointerfaces, and yield new strategies to control the detrimental or beneficial impacts of the microorganisms on the surrounding phases.

Here, we study the response of two different *P. aeruginosa* strains, *P. aeruginosa* PAO1 and PA14, confined at the hexadecane-water interface to address how cell physiology is affected by interfacial confinement. These two strains are ideal model systems because both strains typically form biofilms on solid surfaces, but the composition of their biofilms differs in that PA14 does not secrete Psl polysaccharides. We show that these two strains interact differently with fluid interfaces. We compare the interfacial micromechanics and physiological responses of these two strains and relevant mutants to identify genes essential to elastic film formation at oil-water interfaces.

**Results and Discussion**

We compare the response of *P. aeruginosa* PAO1 and PA14 to hexadecane-water interfaces. To assess bactericidal effects of hexadecane, cells left in hexadecane-saturated media are tested for viability by plating them on LB agar. Both strains survive regardless of the presence of minimal medium supplement (MMS), suggesting hexadecane is not bactericidal (Fig. 1a). In a qualitative demonstration of differing responses, a 5mL volume of microbial suspension in the stationary phase is added to 5mL of hexadecane and vigorously shaken. Droplets in the sample containing PA14 cells are short lived, and the oil and water phases separated rapidly (5 min). However, droplets formed in the presence of PAO1 are highly stable, presumably because of more efficient trapping and restructuring of the interface, allowing oil droplets surrounded by water to remain intact up to 10 days.

To better observe bacteria at oil-water interfaces, we micro-fabricated a platform consisting of 50 μm diameter pores in a PDMS membrane. When this platform is placed on the oil-water interface, bacteria trapped at the interface are visible within each pore. The PA14 cells remain highly motile without forming any apparent structure, while the PAO1 cells accumulate and subsequently restructure the interface to form rigid entities penetrating into the oil phase (Fig. 1b) in the form of a tall, towering, "chef hat"-like structure. These observations establish that the two strains respond very differently to interfaces: PAO1 cells organize into solid-like films at the interface, whereas PA14 cells do not.

**Structure of Films of Bacteria at Interfaces (FBIs).** Films of PAO1 on oil droplets are imaged at different stages of growth (Fig. 1c). Film formation is initiated by individual cells adsorbing and aggregating to fully cover the fluid interfaces. These immature structures are thin films of cells adhered to the interfaces. As these films mature (day 10, (Fig. 1d)), the cells secrete polymer to form interfacial structures. A sample of film recovered from a water-in-hexadecane droplet after 10 days reveal an embedded, interconnected network of cells and secreted polymeric matrix as shown in Figure 1d.

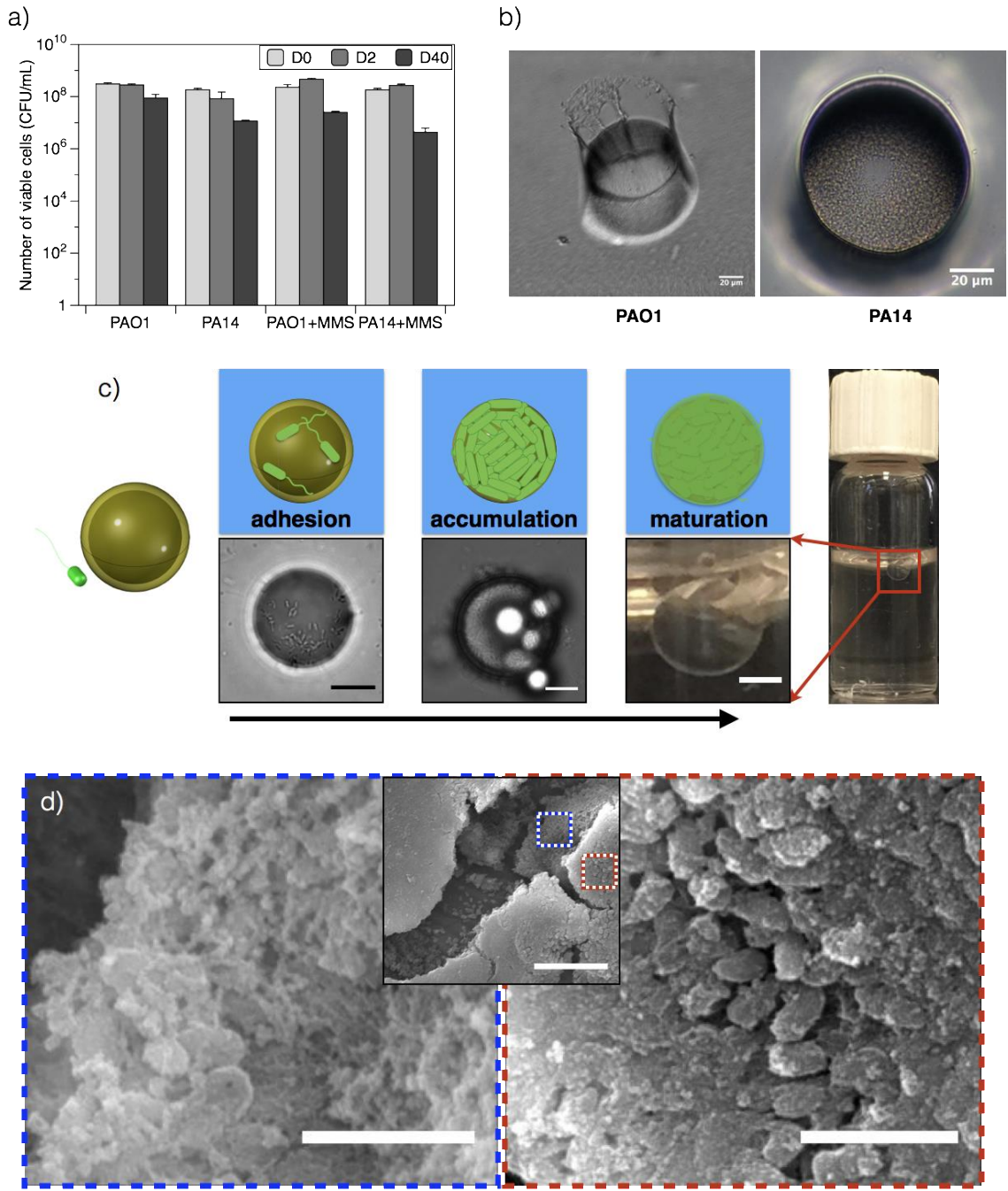

**Figure 1 – FILMS OF BACTERIA AT INTERFACES (a)** Low hexadecane toxicity for PAO1 and PA14 cells. Cells were re-suspended in saline solution (with or without a minimal media supplement, MMS) and exposed to hexadecane for 2 or 40 days and analyzed for viability by counting colony forming units (CFU) on an agar plate after incubation for 24 h. Cells remained viable for 40 days. **(b)** A PDMS platform with 50 microns diameter pores was fabricated to observe PAO1 and PA14 cells confined at the oil-water interface. The cells display a differential response to the interface. PAO1 cells aggregate to form the "chef hat" structure. PA14 cells form an active layer characterized by a highly motile phase (scale bar: 20 µm). **(c)** FBI (Film of Bacteria at Interfaces) formation for *P.*

*aeruginosa* PAO1 cells including the interaction of the bacteria with the oil droplet, adhesion, accumulation of the cells on the oil droplet, and maturation of the film. Single cells adhere to and cover the fluid interface over time. Interfaces are completely covered by a complex structure after 10 days (scale bar: 20 µm, 20 µm and 1mm, respectively). **(d)** Scanning electron microscopic images of interfaces aged for 10 days reveal an asymmetric structure of PAO1 cells within a matrix of extracellular material (scale bar: 10 µm and 2 µm for low- and high-magnification images).

**Mechanical properties of interfaces in contact with PAO1 and PA14 strains.** The macroscopic- and microscopic mechanics of interfaces of bacteria suspensions in contact with hexadecane are characterized via pendant drop elastometry and microrheology, respectively.

In pendant drop elastometry, a small volume of oil is withdrawn from a pendant drop of oil (approximately 5 µL in its original size) that has been aged in a bacterial suspension for 24 hours as shown in in [Fig. 2A (See experimental section for details)](). The drops in contact with the PAO1 suspension show evidence of a wrinkled "bag"-like structure upon compression, suggesting that the FBI covering the interface is a thin solid film with finite bending modulus. A similar experiment for the PA14 suspension, however, shows no such structure formation. Rather, upon compression, the drop shrinks like a typical liquid drop.

To characterize the fluidity of bacteria-laden interfaces, an ensemble of colloidal particles (1 µm in diameter polystyrene particles) is introduced as tracers at the interface and their motion is tracked as a function of interface age for 24 hours. Mean-squared displacements of the ensemble of tracer particles as a function of lag time (evaluated from $1.67 \times 10^{-2}$-1.67 s) obey power laws with exponent *n*, $\langle \Delta r^2(t) \rangle \sim t^n$. An exponent *n*>1, typically observed at early ages, indicates that the particles move super-diffusively, with spatially and temporally extended displacements along paths biased by bacteria motion. An exponent $n$=1 indicates the tracer particles trace random-walk paths; such displacements can occur from repeated interactions with active bacteria or from diffusive Brownian motion; diffusivities inferred from the magnitude of the displacements reveal which is at play. Sub-diffusive motion ($n$<1) indicates a viscoelastic character to the interface. Finally, $n$=0 indicates the entrapment of colloids in a solid, elastic matrix. For a bacteria-free hexadecane-water interface, at the time resolution of our experiments, ~60 fps, particles move diffusively, driven by thermal fluctuation as indicated by $n$=1.

Particle displacements are significantly affected by the presence of bacteria at the interface, and, more importantly, the two bacterial strains show strikingly different behavior (Fig. 2, Panels B1 and B2). For interface ages less than 1800 s, colloids at PAO1-laden interfaces move superdiffusively. However, after 3600 s, the interface changes abruptly to show strong elastic behavior[7]; from this time onward, colloid displacements are highly constrained and subdiffusive. Typical colloid trajectories over 6-second time spans reveal changes in particle displacement as the interface ages (Fig. 2, Panel C1); early in the experiment (at interface ages less than $10^4$ s), colloids traverse distances of tens of micrometers; at later times (at interface age > $3 \times 10^4$ s), the root mean squared displacements (RMS), $d$, at a lag time of 1.67 s, ($\langle \Delta r^2(t = 1.67\ s) \rangle = d^2$), are reduced by more than an order of magnitude (Fig. 2, panel B, open circles) and fall well below that of particles at a bacteria-free hexadecane-water interface. In marked contrast, colloidal probes at interfaces of PA14 suspensions have no evidence of such a transition in their motion even after several hours. These data show unequivocally that PAO1 and PA14 respond differently to the interfacial environment; PAO1 forms an elastic solid film in which embedded colloids are nearly fixed in place, while PA14 shows no evidence of solid elastic film formation. We characterize the elastic properties of the film formed by PAO1 cells at the oil-water interface.[7] As the interface becomes covered with the solid elastic film and *n* approaches 0, the relationship between the interfacial elastic shear modulus, *G'* and asymptotic mean-squared displacement of the colloid $\langle \Delta r^2(t \to \infty) \rangle$ is inversely proportional.

$$G' = \frac{k_B T}{\langle \Delta r^2(t \to \infty) \rangle}$$

Assuming values of *n* < 0.2 are sufficiently elastic, we estimate the modulus of the elastic film to be *G'* ≈3.8 µPa-m after 24 h , similar to elasticity estimated by particle tracking of FBI formed by *Pseudomonas sp.* ATCC 27259[7].

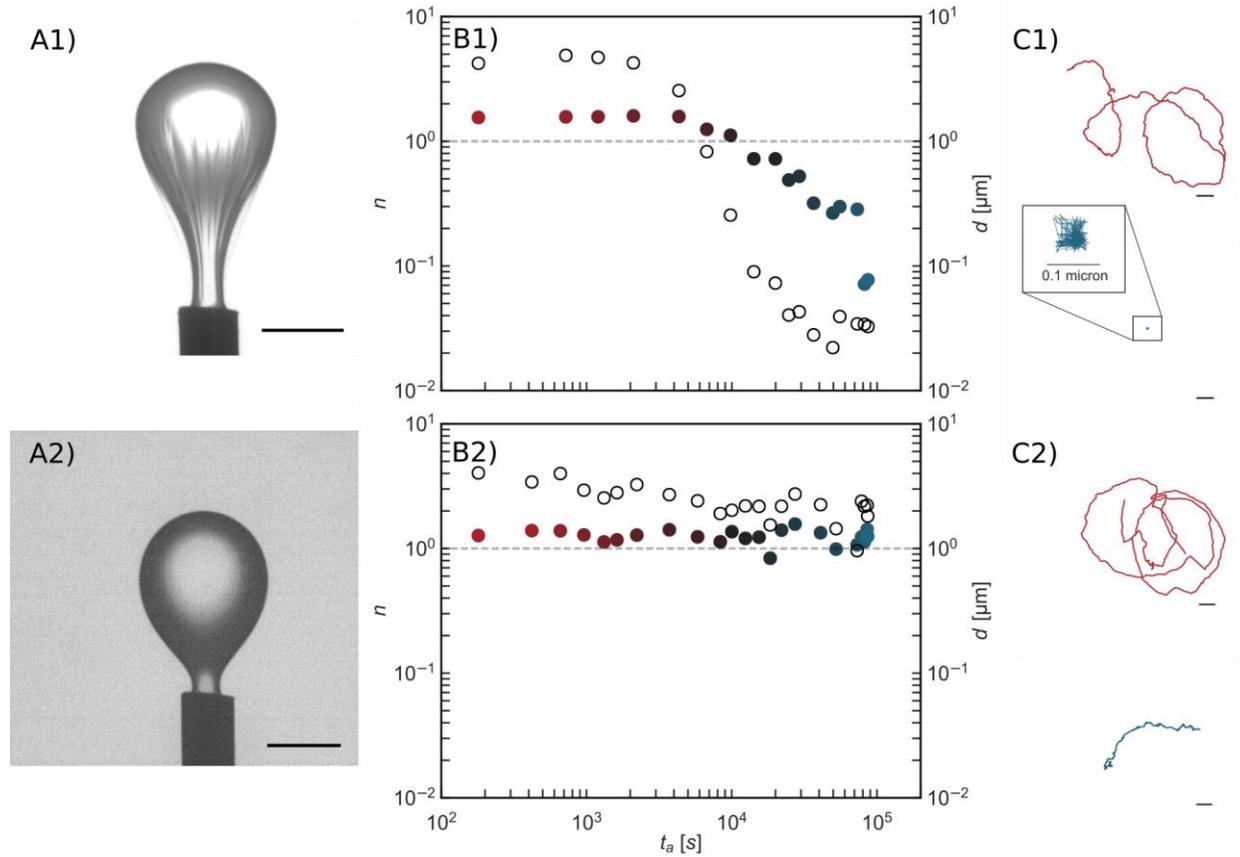

**Figure 2 – CHARACTERIZATION OF THE MECHANICS OF BACTERIA LADEN INTERFACES.** *Panel A*: Pendant drop elastometry on a hexadecane droplet held at the tip of an inverted needle in a bacteria suspension for 24 h and the drop surface is subsequently compressed by withdrawal of hexadecane. Scale bars: 1 mm **(A1)** PAO1 laden interfaces are covered with a solid elastic film after 24 h. **(A2)** PA14 laden interfaces show no evidence of an elastic film formation over the same period. *Panel B*: Exponent $n$ (solid symbols) for lag times from $1.67 \times 10^{-2}$-1.67 s and corresponding root mean square displacements $d$ at lag time of 1.67 s (open circles) versus surface age. Color of the solid symbols, red to blue, indicates increased age. The dashed line, for reference, indicates the values of $n$ and $d$ of colloids subject to thermal Brownian motion in a bacteria-free hexadecane-water interface. **(B1)** PAO1 laden interfaces **(B2)** PA14 laden interfaces. The PAO1 laden interface transitions to a nearly immobile film within $10^4$ s. Probes at PA14-laden interface remain highly mobile for $>10^4$ s. Thereafter, $d$ decays slightly as the interface becomes densely populated with bacteria, but the motion remains diffusive or super-diffusive. *Panel C*: Typical trajectories traced by a colloidal probe for a 6 s time span at early (surface age of 60 seconds) and late ($8.0 \times 10^4$ seconds) interface ages. Scale bars: 1 μm. **(C1)** PAO1 laden interfaces. **(C2)** PA14 laden interfaces. The colloid in the PAO1 laden interface is embedded in an elastic matrix.

**Role of cellular features on elastic film formation.** The different responses of PAO1 and PA14 to the oil-water interface are particularly notable given that both strains form biofilms on solid surfaces. Both PAO1 and PA14 sense and respond to their microenvironment; their motility allows them to swim, swarm, and adopt a collective behavior while assembling into networks and aggregating. We speculate that suppression of the motility functions, or the ability to interact through pili and to secrete polymeric materials might alter their responses to interfacial entrapment and could reveal the contributions of these features to the observed interfacial properties. To explore this idea, we tested the effects of individual deletions of three genes implicated in the formation of biofilms: *flgK*[8], *pilC*[9], and *pelA*[9-11]. The absence of features associated with these genes cause defects in the ability of cells to initiate attachment, form microcolonies, and a robust biofilm matrix. Pendant drop elastometry and microrheology experiments were performed to characterize the interfacial films formed by these mutant strains. The PAO1 mutants, acquired from the Library at the University of Washington, display P1 or P2 phenotypes that are more or less resistant to chloramphenicol, respectively. However, this phenotypic variation did not affect the ability of the PAO1 to form the elastic films at the hexadecane-water interface.

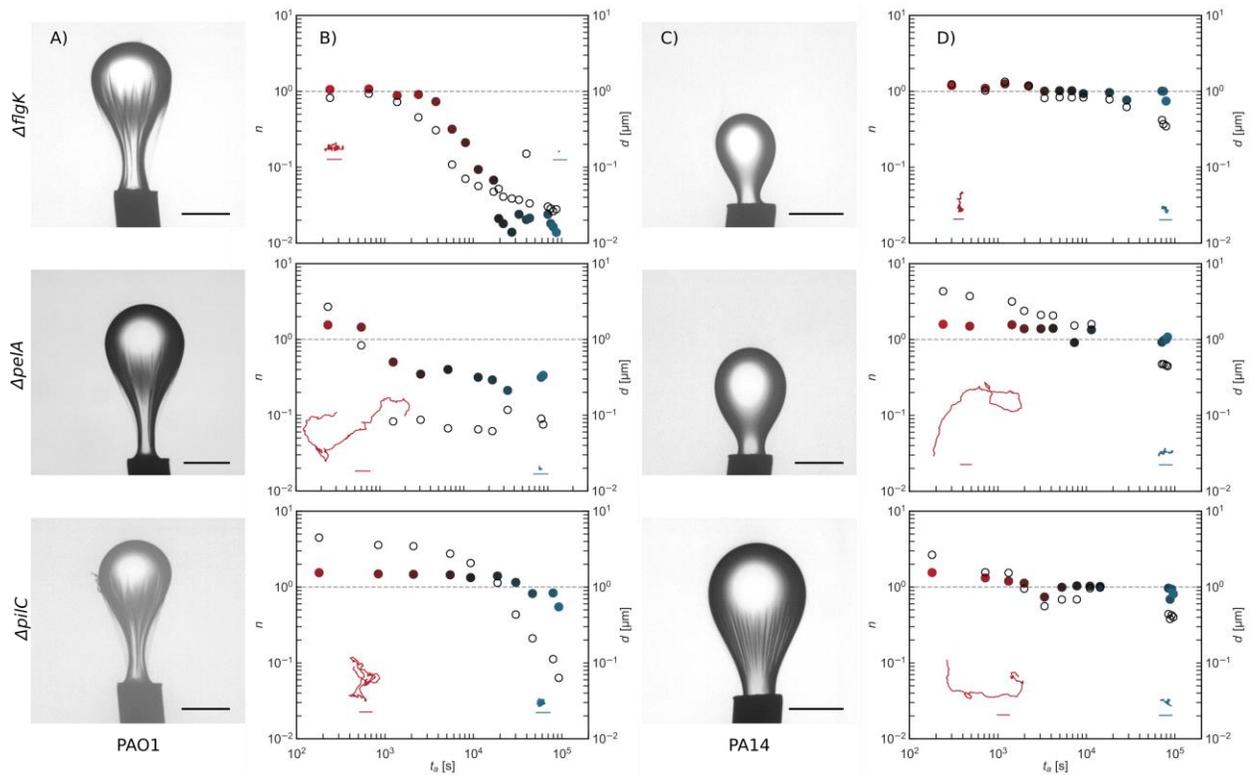

**Figure 3 – MUTANT LADEN INTERFACES**. *Panels A & B*: Pendant drop elastometry, exponent, *n* (solid circles), and root mean square displacement *d* (open cirlces), vs. interface age for PAO1Δ*flgK,* PAO1Δ*pelA,* and PAO1Δ*pilC* laden interfaces. The line at $n = 10^0$ indicates purely diffusive behavior; $n < 10^{-1}$ indicates an essentially solid elastic film. The PAO1Δ*flgK* laden interface transforms from diffusive to elastic layers, while colloids at interfaces in contact with the other mutants (PAO1Δ*pilC*, and PAO1Δ*pelA*) remain superdiffusive and or diffusive indicating that they retain some fluid character over the course of the experiment. All PAO1 knockouts develop skins on pendant drops, evidenced by wrinkles upon compression. Insets to Panel B: Typical trajectories traced by a colloidal probe over 3 s at bacteria laden interfaces at early (~$10^2$ seconds) and late (~$8\times10^4$ seconds) interface ages. Scale bars, 1 micron, are the same for all trajectories. *Panels C & D*: Pendant drop elastometry, exponent, *n*, and *d,* vs. interface age for PA14Δ*flgK,* PA14Δ*pelA,* and PA14Δ*pilC* laden interfaces; the PA14Δ*pilC* alone shows evidence of shear stress supporting film formation on the pendant drop; however, interfaces in contact with these mutants remain superdiffusive or diffusive through very late interface ages. Insets to panel D: Typical trajectories traced by a colloidal probe over 3 s at bacteria laden interfaces at early (~$10^2$ seconds) and late (~$8\times10^4$ seconds) interface ages. Scale bars, 1 micron, are the same for all trajectories.

Remarkably, all three mutants (ΔflgK, ΔpilC, and ΔpelA) of PAO1 display the phenotype of a wrinkled elastic "bag" similar to that observed in the pendant drop experiments with the wild-type PAO1 (Fig. 3, Panel A). However, the films formed by PAO1 mutants (Fig. 3, Panel B) have noticeable changes in dynamic response, suggesting different roles of each cellular feature on the formation of interfacial film. The suppression of motility functions in the PAO1ΔflgK cells alters the early dynamics at the interface; colloidal probes move diffusively with displacements like those provided by Brownian motion ($d(1.67\ s) \approx 1\mu m$). The interface is covered by a purely elastic film, which forms within ~$10^4$ sec, with a modulus of 1 µPa.m, which increased monotonically to 6 µPa.m after 24 h. The mutants PAO1ΔpelA and PAO1ΔpilC retain their flagella but lack the ability to produce extracellular polymer matrix and pili, respectively. The films formed by the PAO1ΔpelA mutant transition from a superdiffusive regime to a subdiffusive regime after only $10^3$ sec. The films formed by the de-piliated bacteria transition to subdiffusive much later. For both films, even after 24 h, the films formed by these mutants have exponent *n* of 0.4 and 0.5, suggesting that interfaces covered with these mutants are not fully elastic but retain a viscoelastic character at this age. These results suggest that pili and secreted molecules play some role in the formation of elastic films at the oil-water interface. However, their removal does not completely eliminate the formation of a skin with elastic characteristics, as seen in the wrinkles of the compressed film on the surface of the pendant drops.

All three PA14 mutants form active layers in which colloids move first superdiffusively, then diffusively, with slightly subdiffusive motion only at very late interface ages (Fig. 3, Panels C and D). Colloid motion is slowed for these interfaces at interface ages > $10^4$ s, attributable to crowding at the interface as bacteria proliferate. Compression of pendant drops aged in PA14Δ*flgK* and PA14Δ*pelA* suspensions gave no evidence of elastic film formation, while those in contact with suspensions of PA14Δ*pilC* cells show characteristics of weakly viscoelastic films, only at late interface ages (~$10^4$ s). Furthermore, upon withdrawal of oil from the pendant drop aged in contact with PA14Δ*pilC*, wrinkles form, suggesting a finite bending modulus to this layer. Interestingly, films of these de-piliated microbes also show evidence of microbes leaving the interfacial region upon compression. Upon compression, a dark plume, presumably of bacteria ejected from the film, is observed adjacent to the pendant drop, reminiscent of desorbed plumes of nanoparticles from a compressed pendant drop as reported previously[12]. Some bacteria escape from the film formed by the PA14Δ*pilC* mutants upon drop compression from the apex of the drop, indicating that some population is only weakly cohered to the film. These cells may also rearrange within the interface, evidenced by the gradual disappearance of the compression-induced wrinkles without a change in pendant drop volume. Colloids at interfaces of suspensions of PA14Δ*pelA* cells (which retain both flagella and pili) trace superdiffusive paths longer than the de-piliated counterparts, and move diffusively at late interface ages. The interpretation of random colloidal displacements in a medium that contains active elements, including living entities, must be handled with care [13,14]. Diffusive trajectories in a fluid indicate random interactions with active elements when the colloid diffuses faster than thermal motion would imply. However, random displacements can also occur for particles trapped in a viscoelastic matrix undergoing stochastic forcing., The absence of apparent elastic structure at the PA14-laden interfaces in the pendant drop experiments suggests that PA14-laden interfaces remain viscous films. The main findings of this work regarding the major differences in structure formation between PAO1 and PA14 laden interfaces do not rely on these details.

Our results indicate that PAO1 cells and their mutants form FBI even in the absence of cellular functions known to be important to biofilm formation on solid substrates. These findings suggest that while some cellular features such as secreted polysaccharides and pili contribute to the elasticity of the FBIs, they are not the most important features that determine the formation of elastic films at the interface.

**Cell viability at fluid interfaces.** In addition to characterizing the effect of gene deletion (Δ*flgK*, Δ*pilC*, and Δ*pelA*) on the mechanics of films of bacteria at the interface, we studied mutants lacking other functions relevant to biofilms, such as the genes controlling the ability to secrete polysaccharide *pslD*, and rhamnolipid *rhlA*. In none of these cases did a PAO1 mutant fail to form a film that had elastic character at the interface. Only the addition of 100μg/mL Ciprofloxacin, a bactericidal antibiotic, to the suspension of the PAO1 wild type suppresses the formation of these interfacial structures, which leads us to question if the formation of an elastic interfacial film occurs as a dynamic response to ensure survival. Can the interfacial films provide a structural but also protective function to the embedded cells to ensure survival? To test this hypothesis, we determine the viability of stationary PAO1 and PA14 cells adhered to a hexadecane-water interface for 1 h. Droplets of hexadecane are exposed to the bacterial suspensions and gyratory rotation (60 rpm) is applied for 1h to allow the droplet to stabilize. The emulsions (oil in water) are stained using a live/dead assay (Filmtracer LIVE/DEAD Biofilm Viability Kit, Thermo Fisher Scientific) to evaluate the number of live cells at the interfaces. The imaging of the droplet interfaces indicates that PAO1 cells at fluid interfaces remain viable over the duration of the experiments. In the case of PA14, most cells at the interface also remain viable although we observed a small number of droplets with dead cells. These findings suggest that both cells are able to survive the interfacial confinement but respond very differently. To better understand the physiological stresses that the cell experiences, we analyze their transcriptional profile.

**Transcriptional profile of cells associated with active layers and elastic films.** Transcriptional profiles of cells trapped at the oil-water interface for 1 h were obtained via RNA Seq in triplicate experiments. Gene expression greater than a 2-fold change and false discovery rate (FDR)<0.01 were considered significant. The change in the transcriptomes of the cells confined at the hexadecane-water interface in comparison to the control is described in Figure 4. Genes significantly upregulated and downregulated by interaction with the interface (vs. an untreated control, a planktonic suspension in a 154 mM NaCl solution) are highlighted in blue and red, respectively (Fig. 4). The transcriptomes of the *P. aeruginosa* cells in the early stage of the interfacial confinement reveal that PA14 cells show a larger number of expression levels that are significantly up- or down-regulated compared to that by PAO1. The functions of the top 10 genes induced or repressed in PAO1 or PA14 cells listed in Table 1 & 2 differ strongly, suggesting that distinct pathways may be involved in the response of PAO1 compared to PA14.

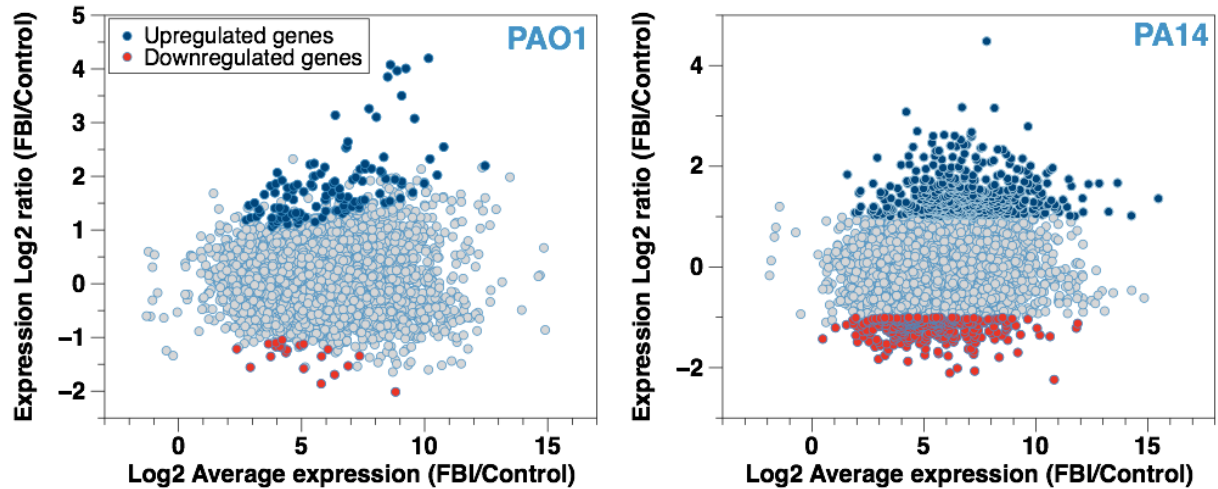

**Figure 4 – MA (MEAN-AVERAGE) PLOT FOR GENE EXPRESSION UNDER FBI CONDITIONS**. Log2-fold of the change in expression vs. the Log2 average expression for each gene expressed by (a) PAO1 and (b) PA14 cells after confinement at a hexadecane-water interface for 1 h is presented. The total number of total mRNA expressed by the cells was plotted as a function of expression ratio. The blue and the red dots represent genes with significant changes in expression, by at least 2-fold with a false discovery rate (FDR)<0.01.

Interfacial trapping of the PA14 cells induces major transcriptional changes compared to the control cells not exposed to the interface. A total of 482 and 278 PA14 genes involved in PA14 active layers are either induced or repressed, respectively. The most expressed functions, besides the *hypothetical functions*, include those categorized as *transport of small molecules*, *putative enzymes,* and *transcriptional regulators*. PA14 cells respond to the interfacial confinement through the transcription of genes with functions including *putative enzymes* (25 downregulated, 38 upregulated), *two-component regulatory systems* (7 downregulated, 11 upregulated), and other *transcriptional regulators* (35 downregulated, 27 upregulated). Genes related to *cell wall, LPS, and capsule* (5 downregulated, 6 upregulated), and *membrane proteins* (16 downregulated, 18 upregulated) also show changes in PA14. A total of 13 motility genes (3 downregulated, 10 upregulated) contribute to the active behavior of the PA14 cells at the interface. In addition, 10 genes (8 upregulated and 2 downregulated) involved in *DNA replication, recombination, modification, and repair* are significantly expressed under the conditions of interfacial confinement, along with genes involved in *energy metabolism* (8 downregulated vs. 22 upregulated genes) and *chemotaxis* (8 upregulated vs. 2 downregulated genes). While the increased transcription of *DNA repair* genes could reflect an increase in DNA replication, its change in expression could also imply that the PA14 cells experience physiological stress at the oil-water interface causing DNA damage[15-17].

Table 1: Top 10 genes induced or repressed in PAO1 FBIs vs. control

| PAO1 Gene | Description | Functions | Fold Change |
|---|---|---|---|
| *Downregulated* | | | |
| PA3920 | probable metal transporting P-type ATPase | Transport of small molecules | -4.0 |
| PA3574 | NalD | Transcriptional regulators | -3.6 |
| PA3520 | hypothetical protein | Hypothetical, unclassified, unknown | -3.2 |
| phzC2 | phenazine biosynthesis protein PhzC | Secreted Factors (toxins, enzymes, alginate) | -3.0 |
| PA0883 | probable acyl-CoA lyase beta chain | Putative enzymes | -2.9 |
| PA4220 | hypothetical protein | Hypothetical, unclassified, unknown | -2.9 |
| PA3207 | hypothetical protein | Hypothetical, unclassified, unknown | -2.6 |
| PA3330 | probable short chain dehydrogenase | Putative enzymes | -2.6 |
| PA2330 | hypothetical protein | Hypothetical, unclassified, unknown | -2.5 |
| mdcE | malonate decarboxylase gamma subunit | Carbon compound catabolism | -2.4 |
| *Upregulated* | | | |
| PA4682 | hypothetical protein | Hypothetical, unclassified, unknown | 18.4 |
| PA0830 | hypothetical protein | Hypothetical, unclassified, unknown | 16.9 |
| PA1538 | probable flavin-containing monooxygenase | Putative enzymes | 16.1 |
| PA3427 | probable short-chain dehydrogenases | Putative enzymes | 15.6 |
| PA2550 | probable acyl-CoA dehydrogenase | Putative enzymes | 14.4 |
| alkB2 | alkane-1-monooxygenase 2 | Carbon compound catabolism | 11.3 |
| PA1542 | hypothetical protein | Hypothetical, unclassified, unknown | 9.6 |
| PA3323 | conserved hypothetical protein | Hypothetical, unclassified, unknown | 8.8 |
| PA3277 | probable short-chain dehydrogenase | Putative enzymes | 8.6 |
| fliE | flagellar hook-basal body complex protein FliE | Motility & Attachment | 8.4 |

Table 2: Top 10 genes induced or repressed in PA14 FBIs vs control

| PA14 Gene | Description | Functions | Fold Change |
|---|---|---|---|
| *Downregulated* | | | |
| PA14_28410 | hypothetical protein | Hypothetical, unclassified, unknown | -4.7 |
| PA14_33830 | hypothetical protein | Hypothetical, unclassified, unknown | -4.3 |
| PA14_09450 | phenazine biosynthesis protein PhzD | Secreted Factors (toxins, enzymes, alginate) | -4.2 |
| PA14_21520 | hypothetical protein | Biosynthesis of cofactors, prosthetic groups and carriers | -4.2 |
| PA14_33520 | Thioesterase | Biosynthesis of cofactors, prosthetic groups and carriers | -4.0 |
| PA14_10490 | hypothetical protein | Hypothetical, unclassified, unknown | -3.7 |
| PA14_11330 | hypothetical protein | Hypothetical, unclassified, unknown | -3.6 |
| PA14_03090 | hypothetical protein | Hypothetical, unclassified, unknown | -3.5 |
| PA14_03080 | Acetyltransferase | Transport of small molecules | -3.4 |
| PA14_10160 | ferric enterobactin transport protein FepD | Transport of small molecules | -3.4 |
| *Upregulated* | | | |
| PA14_19730 | Oxidoreductase | Putative enzymes | 22.4 |
| PA14_21640 | short chain dehydrogenase | Biosynthesis of cofactors, prosthetic groups and carriers | 9.0 |
| PA14_19700 | Aldolase | Carbon compound catabolism | 8.9 |
| PA14_08230 | hypothetical protein | Hypothetical, unclassified, unknown | 8.5 |
| PA14_46880 | glutathione synthase | Putative enzymes | 6.9 |
| PA14_08020 | bacteriophage protein | Related to phage, transposon, or plasmid | 6.5 |
| PA14_31580 | acyl-CoA dehydrogenase | Fatty acid and phospholipid metabolism | 6.4 |
| PA14_08240 | hypothetical protein | Related to phage, transposon, or plasmid | 6.2 |
| PA14_08200 | hypothetical protein | Hypothetical, unclassified, unknown | 6.2 |
| PA14_43370 | potassium-transporting ATPase subunit C | Transport of small molecules | 6.1 |

Confinement at the fluid interfaces results in relatively fewer changes in the transcriptional profile of PAO1 cells. In total, 118 genes and 21 genes are induced and repressed, respectively, most of which are annotated as *hypothetical function*. A closer examination of the genes of known function indicates that the PAO1 cells did not show significant expression changes for any *DNA repair* genes, which suggests that the elastic interfacial film might protect the cells from deleterious effects of the interfacial environment. Instead, functions such as *motility* (5 upregulated, 1 downregulated genes), *carbohydrate metabolism* (2 upregulated, 2 downregulated genes), *adaptation and protection* (2 upregulated genes), *cell wall, membrane and envelope biogenesis* (5 upregulated genes) are implicated in the transcriptional response of PAO1 to interfacial confinement for 1 h. Also, 19 genes *related to phage, transposon, or plasmid* are upregulated under interfacial confinement. The implications of these functions in PAO1 response to interfacial confinement remain to be elucidated.

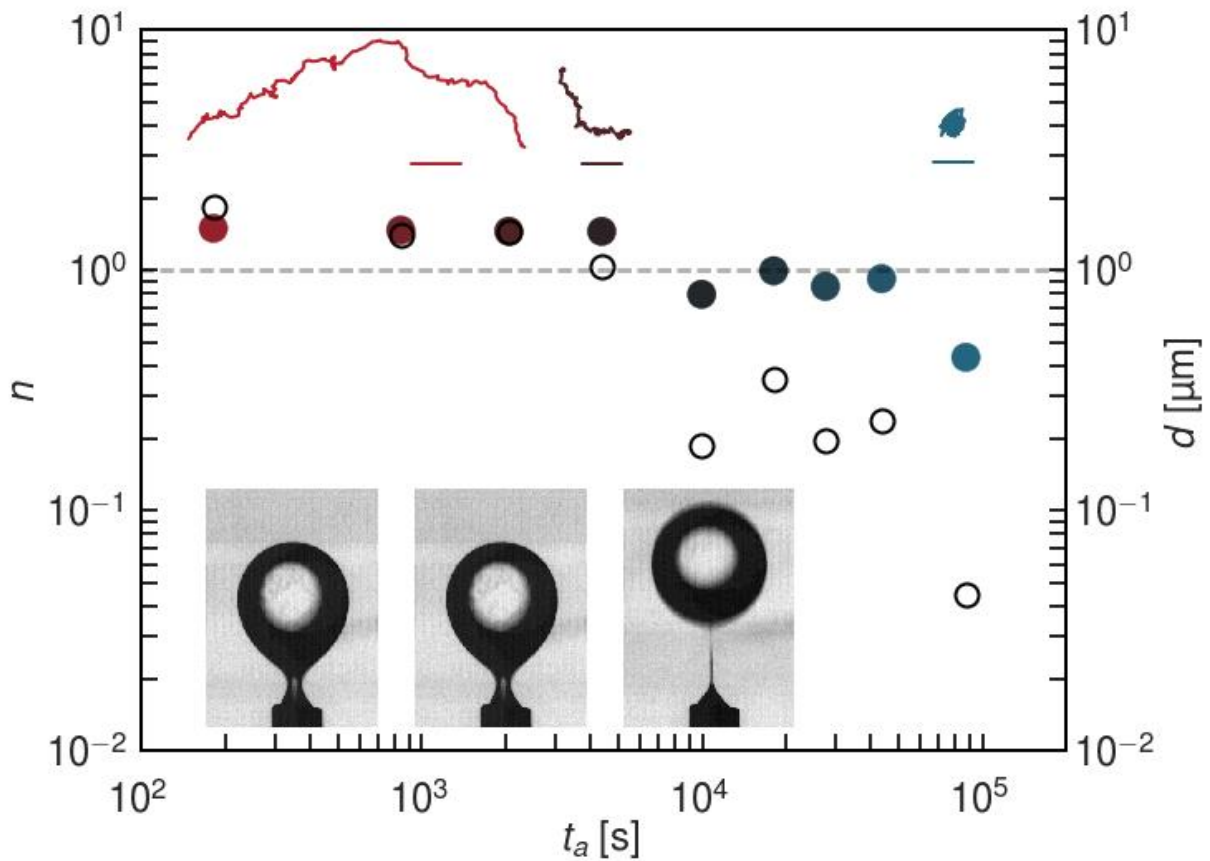

**Figure 5 – MECHANICS OF PAO1ΔalkB2 LADEN INTERFACES.** Pendant drop compression (from left to right) performed on a droplet of hexadecane held in a suspension of PAO1Δ*alkB2* for 24 h shows no evidence of elastic film formation; surface tension decreased until drop detached. Colloidal displacements remain superdiffusive for hours (lag times up to 1.67 s); the exponent $n$ and RMS $d$ only decay after $10^4$ s; probe motion at late ages (~$8\times10^4$ s)

is significantly slowed and slightly subdiffusive, suggesting that the colloids are embedded in a weakly viscoelastic fluid interface rather than in a solid elastic film.

Of particular interest are 10 genes encoding for *putative enzymes* that are upregulated, including 6 genes (*PA1538, PA3427, PA2550, PA3277, PA1648, PA0840*) induced 6- to 18-fold. Among these upregulated genes, *PA1538* is induced 16-fold and encodes a probable flavin-containing monooxygenase that may play a role in oxidizing, breaking down, and expelling xenobiotics or foreign materials[18]. In addition, the *alkB2* gene, which has been shown to be an alkane hydroxylase involved in $C_{12}$-$C_{16}$ alkane oxidation and catabolism[19,20], is induced by 11.3 fold under interfacial exposure, suggesting that PAO1 might cope with the interfacial stress by metabolizing hexadecane. The upregulation of this gene, and possibly other genes for *putative enzymes*, may suggest the hexadecane-water interface is conducive to cell growth[19,20] and/or adaptation through hexadecane modification.

To study the role of *alkB2* in PAO1's elastic film formation at interfaces, mutant cells lacking the gene were trapped at the hexadecane-water interface. Surprisingly, PAO1Δ*alkB2* is unable to form an elastic film after 24 h. This is evident in the pendant drop experiment (Fig. 5), and further confirmed by microrheology. Thus, deletion of *alkB2* in PAO1 suppresses the evolution from active to elastic regime commonly displayed by the wild type, indicating that hexadecane metabolism through *alkB2* expression contributes to the formation of PAO1's elastic interfacial film. More intriguingly, while the *alkB2* gene is present in PA14 (*PA14_44700*), its expression is not significantly upregulated, suggesting that the response to interfacial confinement of the two strains may involve significantly different metabolic pathways. Further analysis of the genes identified in the transcriptome analysis will allow us to determine if the ability to resist interfacial stress and/or hexadecane metabolism play a key role in the differential behaviors of the PAO1 vs. PA14 cells during hexadecane-water entrapment.

**Conclusion**

Bacteria can form elastic films at fluid interfaces by a dynamic process that occurs as a response by bacteria to interfacial entrapment. In this study, we have compared the response of two *P. aeruginosa* strains to entrapment at the hexadecane-water interface. PAO1 cells display a dramatic transition from active and superdiffusive to elastic states at the oil-water interface, in which bacteria first become trapped at the interface in a motile layer, and subsequently restructure the interface to form an elastic film. However, the PA14 cells do not form an elastic film, but rather remain motile over extended periods of time. Transcriptional profiles demonstrate that in PAO1, very few genes are expressed in the early interactions of the bacterium with the interface. However, genes that may play roles in hexadecane metabolism are induced. This may have direct implications for forming an elastic film, e.g. through chemical modification of the hexadecane, or indirect implications, as cells may exploit the alkane as a carbon and energy source. In contrast, the initial response of PA14 cells to adopt a monotonic behavior may indicate an inability, or impaired ability, to metabolize the hexadecane and generate a protective matrix. These results may help to define new metabolic pathways to improve the hydrocarbonoclastic abilities of microorganisms, and to develop new genetic targets to prevent the deterioration of economically relevant products.

**Materials and Methods**

***Microorganisms and growth condition:*** *P. aeruginosa* PAO1 and PA14 cells were used in all the experiments. The corresponding mutants for PAO1, including PW3751 (*alkB2*-A07::ISlacZ/hah), PW1664 (*flgK*-A12::ISlacZ/hah), PW8626 (*pilC*-B07::ISphoA/hah), PW6140 (*pelA*-H06::ISphoA/hah), PW4802 (*pslD*-H08::ISlacZ/hah), and PW6886 (*rhlA*-E08::ISphoA/hah), were acquired from a mutant library[21]. The cells were grown in Lysogeny broth (LB) composed of 10g/mL tryptone, 10 g/mL NaCl, and 5 g/mL yeast extract resuspended in Milli-Q water water. The experiments were conducted with cells grown at 37°C for 16-18 h in 25 mL LB medium. Following culture overnight, the cells were collected in a 50 mL conical tubes, centrifuged and washed three times with 154mM NaCl solution. For all the experiments, the cells were finally diluted in 154 mM NaCl solution to reach an optical density at 600 nm of 0.2.

***Minimal Medium Supplement (MMS):*** A liter of MMS included L-Ascorbic acid (0.035 g), Sodium acetate (0.07 g), Sodium thiosulfate (0.1 g), Sodium nitrate (0.12 g), Succinic acid (0.37 g), L-Tartaric acid (0.37 g), Potassium phosphate monobasic (0.68 g), 5 ml Wolfe's Mineral solution, and 5 ml Wolfe's vitamin solution in addition to milli-Q water.

***Particle-tracking:*** As described previously[7], particle-tracking is performed by spreading polystyrene microspheres (Invitrogen) with radius $R_s$ = 0.5 μm at a planar interface formed between the bacteria suspension and hexadecane. This planar oil-water interface is created by first pinning a layer of bacteria suspension in a 1 cm inner diameter cylinder whose bottom half is aluminum and top is Teflon. The aqueous suspension is placed in the bottom half of this vessel, with interface pinned at the height where aluminum and Teflon intersect; the volume

(~450 µL) is adjusted to insure a flat interface. A 1 µL drop of microspheres in a spreading solution containing isopropyl alcohol, ethanol, and water is placed in contact with the bacteria suspension.  The microspheres and spreading solution rapidly spread on the interface. Hexadecane is gently pipetted on top of the bacteria and microsphere-laden interface to fill the cylinder.  The interface age is calculated from the time that this oil-aqueous interface is formed. Videos of the particle motion, taken at 60 fps, are analyzed using a custom Python implementation[22] of the Crocker-Grier multiple-particle-tracking algorithm[23] to locate regions of high intensity. The particle locations are filtered by intensity and feature size in pixels, as well as eccentricity. Locations between frames are linked based on distance a particle is likely to travel to determine the displacement of each particle. The systematic drift is calculated as the average motion in x and y directions and subtracted from the displacements frame to frame, upon which point the ensemble statistics for 25 to upwards of 200 trajectories per interface age, such as mean squared displacement, are calculated.

**Pendant drop Elastometry:** A cuvette is filled with the 154 mM NaCl suspension of the relevant bacteria strain. A microliter-sized droplet of hexadecane is formed and held at the tip of an inverted needle attached to a hexadecane filled syringe. The drop is aged in bacteria suspension for 24 ± 2 h. An image of the drop is captured on a CCD camera. The aged drop is compressed by withdrawing oil via a syringe pump at a constant flowrate of 0.1 ml per minute. As this occurs, the interfacial film tension decreases. If the drop interface remains in a fluid state, the drop shapes remain solutions of the Young Laplace equation as oil is withdrawn. However, is an FBI is present, a wrinkled skin forms on the droplet. The appearance of these wrinkles and the departure of the drop shape from shapes corresponding to solutions of the Young Laplace equation are diagnostic of a thin solid film at the interface.

In all experiments, cell densities were held constant at $(2.97\pm0.38)\times10^9$ CFU/mL. Pendant drop elastometry was used to infer the presence of a solid film. A pendant drop of oil formed in bacteria suspension was aged for 24 hours; a small volume (~ 5 µL) of oil was withdrawn and images of the drop were recorded. The appearance of a wrinkled film wrapping the oil droplet indicates that the interface is covered with a compressed thin solid film with finite bending modulus[7]. Microrheology was used to characterize the film viscoelastic properties. The motion of colloidal polystyrene tracers introduced at the interface was recorded at various interface ages over a 24 hours timespan. Mean-squared displacements (MSDs) of individual tracer particles for lag times up to ~2 s were ensemble averaged; these ensemble averaged MSD obey power laws with exponent *n* that reveals viscoelastic properties. For a hexadecane-water interface absent bacteria, at the time resolution of our experiments, particles move diffusively by thermal motion, and *n*=1. With bacteria, particle displacements at early interface ages are dominated by encounters with the microbes. For large enough lag times, particles in bacterial suspensions or at interfaces typically move diffusively in a viscous environment (with n=1) with an apparent diffusion coefficient initially well in excess of thermal diffusion, indicating the greater kinetic energies imparted to the colloids via interaction with microbes[24,25]. However, at lag times small compared to characteristic relaxation times of the system, colloids move with *n*>1, exhibiting super-diffusive motion, i.e. that particles move in preferred directions[7,26]. If viscoelastic structures form in the interface, colloidal displacement is reduced, and the exponent decreases, *n*<1. For a purely elastic film, *n*=0.

***Electron Microscopy:*** SEM was performed to examine architectures of mature films of PAO1 cells. The PAO1 FBI was formed for 10 days at the hexadecane-water interfaces. The samples were gently rinsed in PBS and fixed with 2.5% glutaraldehyde and 2.0% paraformaldehyde in 0.1M cacodylate buffer (pH 7.4) for 1 hour at room temperature. Finally, the samples were imaged under environmental conditions (low vaccum, 0.75 Torr) using a FEI Quanta 600 ESEM high vacuum Scanning Electron Microscope (Oregon, USA) at an accelerating voltage of 15kV.

***RNA isolation and sequencing:*** To evaluate changes in the transcriptional profile of cells after interfacial confinement, PAO1 and PA14 cells were exposed to the hexadecane-water interface and the cell total RNAs were isolated. First the cells were grown for 16 h in the LB medium, washed and resuspended in 154 mM NaCl solution as described above. To form the FBIs, 5 mL of suspensions and 5mL of hexadecane were introduced in 25 mL glass vials. The vials were rotated at 60 rpm for 2h using a Rugged Rotator (Glas-Col, Terre Haute, USA) for 1 h and the emulsions were collected in microcentrifuge tubes, then centrifuges at 15,000 g for 2.5 min at 4°C to remove cells, which are not attached to the FBIs. The FBIs were collected from the hexadecane water interfaces, introduced in a vial and store at -80°C for RNA isolation. Similarly the control cells, which were only suspended in 154 mM NaCl solution, were collected by centrifugation and store -80°C for RNA extraction. Three independent samples were prepared for the FBIs the control experiments. The RNeasy Mini Kit (Qiagen, CA, USA), which includes on-column DNase, was employed to extract the total RNA of cells following a protocol previously developed[27,28]. Finally, the 3 controls and the 3 FBI samples for PAO1 and PA14 were sent to the Penn next generation sequencing core facility to perform RNA sequencing. The total RNA of PA14 and PAO1 FBIs as well as the untreated control (PAO1 and PA14 in suspension) were isolated and bioanalyzed for their quality. The RNA samples were sent to the *Penn Next-Generation Sequencing Core* to acquire the transcriptional profile of the cells using RNA sequencing.

**Author Contributions**

T. H. R. N., L.V, D. L., and K. S. conceived the experiments, T. H. R. N. and L. V. conducted the experiments, all the authors have analyzed and discussed the results, D. L., K. J. S., R. L. L., and M. G. supervised the work. The manuscript was written through contributions of all authors. All authors have given approval to the final version of the manuscript.

**Acknowledgments**

This research was made possible in part by a grant from The Gulf of Mexico Research Initiative, NSF Grant Nos. DMR-1120901 (Penn MRSEC) and DMR-1610875, NIH Grant No. GM080279 (M.G.), PIRE-1545884. T. H. R. N. was supported by the Postdoctoral Fellowship for Academic Diversity Program (University of Pennsylvania). We thank Knut Drescher (Max Planck Institute for Terrestrial Microbiology, Germany) for all *P. aeruginosa* PA14 strains. Data are publicly available through the Gulf of Mexico Research Initiative Information & Data Cooperative (GRIIDC) at https://data.gulfresearchinitiative.org (doi: 10.7266/N70V89WF, doi: 10.7266/N7W37TFX).

**Conflict of interest**

The authors declare no conflict of interest.